\documentclass[sigchi]{acmart}
\makeatletter
\renewcommand\@formatdoi[1]{\ignorespaces}
\makeatother

\AtBeginDocument{%
  \providecommand\BibTeX{{%
    \normalfont B\kern-0.5em{\scshape i\kern-0.25em b}\kern-0.8em\TeX}}}

\setcopyright{rightsretained}
\copyrightyear{2019}
\acmYear{2019}
\acmDOI{10.1145/1122445.1122456}

\acmConference[DLG@KDD '19]{Proceedings of DLG@KDD workshop (DLG@KDD’19)}{August 2019}{Anchorage, Alaska, US}
\acmBooktitle{{Proceedings of DLG@KDD workshop (DLG@KDD '19) }}
\acmPrice{15.00}
\acmISBN{978-1-4503-9999-9/18/06}

\usepackage{todonotes}

\begin{document}

\title{ncRNA Classification with Graph Convolutional Networks}

\author{Emanuele Rossi}
\affiliation{%
  \institution{University of Cambridge}
}
\email{er513@cam.ac.uk}

\author{Federico Monti}
\affiliation{%
  \institution{USI Lugano}
}
\email{federico.monti@usi.ch}

\author{Michael Bronstein}
\affiliation{%
  \institution{Imperial College London}
  \institution{USI Lugano}
  %\institution{Fabula AI}
}
\email{m.bronstein@imperial.ac.uk}

\author{Pietro li\`{o}}
\affiliation{%
 \institution{University of Cambridge}
}
\email{pl219@cam.ac.uk}

\renewcommand{\shortauthors}{Rossi, et al.}

\begin{abstract}
% Non-coding RNA (ncRNA) are RNA sequences which don't code for a gene but instead carry important biological functions. The task of ncRNA classification consists in classifying a given ncRNA sequence into its family. While it has been shown that the graph structure of an ncRNA sequence folding is of great importance for the prediction of its family, current methods make use of machine learning classifiers on hand-crafted graph features. We improve on the state-of-the-art on this task with a graph convolutional network model which is able to differentiate between bond types. Moreover, our model learns in an end-to-end fashion from the raw RNA graphs and removes the need for expensive features extraction. To the best of our knowledge, this also represents the first successful application of graph convolutional networks to RNA data.

Non-coding RNA (ncRNA) are RNA sequences which don't code for a gene but instead carry important biological functions. The task of ncRNA classification consists in classifying a given ncRNA sequence into its family. While it has been shown that the graph structure of an ncRNA sequence folding is of great importance for the prediction of its family, current methods make use of machine learning classifiers on hand-crafted graph features. We improve on the state-of-the-art for this task with a graph convolutional network model which achieves an accuracy of 85.73\% and an F1-score of 85.61\% over 13 classes. Moreover, our model learns in an end-to-end fashion from the raw RNA graphs and removes the need for expensive feature extraction. To the best of our knowledge, this also represents the first successful application of graph convolutional networks to RNA folding data.
\end{abstract}




\maketitle

\section{Introduction and Related Work}

\begin{figure}[h]
    \centering
    \includegraphics[width=0.5\textwidth]{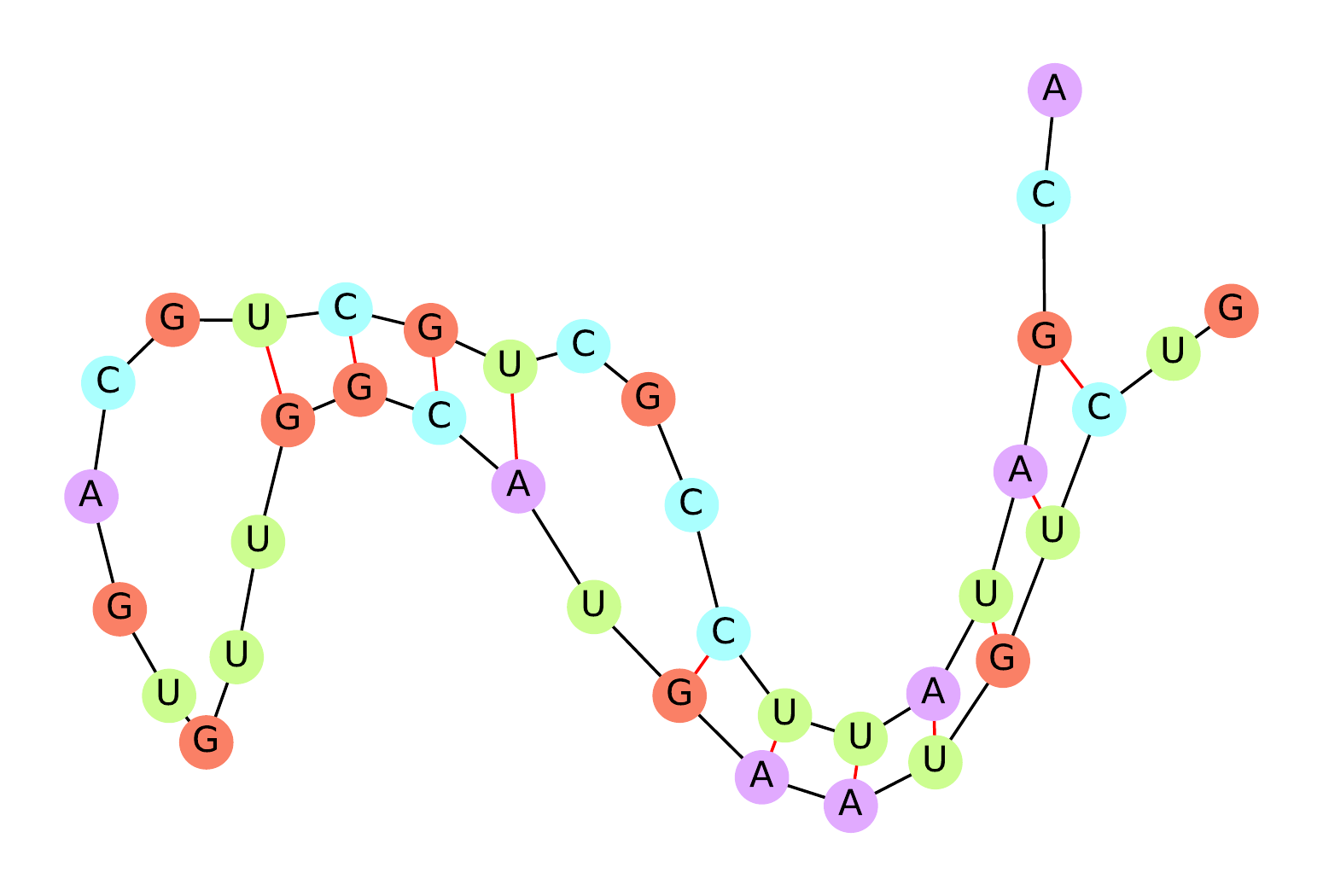}
    \caption{Graph representing the folding of RNA sequence \textit{ACGAUAUUCCGCUGCUGCAGUGUUGGCAUGAAUGUCUG}. Generated using the ViennaRNA package \cite{Hofacker2003}. Black edges represent phosphodiester bonds and red edges represent hydrogen bonds.}
    \label{fig:rna_graph}
\end{figure}

\begin{figure*}[t]
    \centering
    \includegraphics[width=1\textwidth]{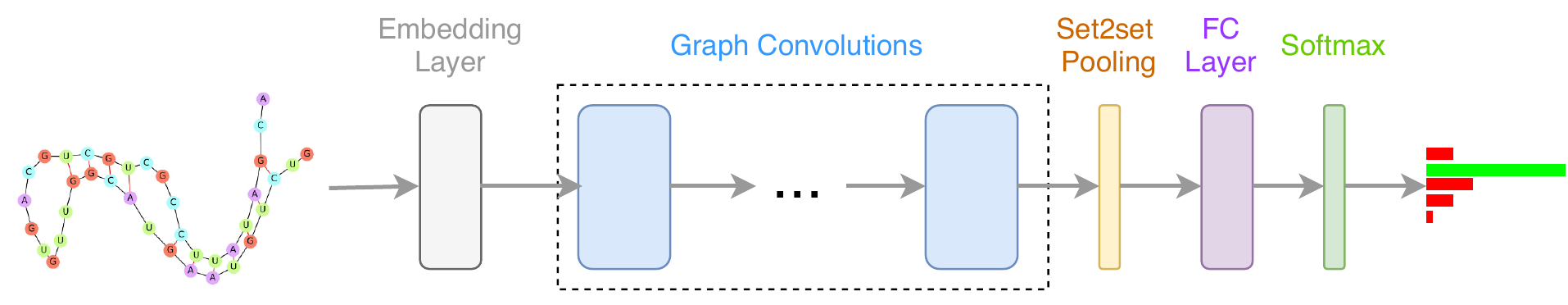}
    \caption{Diagram representing the architecture of our \textit{RNAGCN} model.
    The input consists of an RNA graph with a nucleotide on each node. Initially, an embedding layer maps each nucleotide into a continuous vector. After that, multiple graph convolutional layers are used to refine the features on each node by propagating information in the graph. The set2set pooling model is then used to aggregate the relevant information from the output of the last convolutional layer and produce a graph-wise representation. A fully-connected layer with softmax activation finally produces the output probability for each class.}
    \label{fig:model}
\end{figure*}
    
% An embedding layer is followed by a sequence of graph convolutional layers and the set2set model for global pooling. A fully connected layer followed by a softmax produces then the output probability for each class.

% Taking as input an RNA graph, the model first makes use of an embedding layer to embed the nucleotide on each node into a fixed-length vector. The embedding layer is followed by a sequence of graph convolutional layers, a pooling using the set2set model to produce a representation for the whole graph, and then a fully connected layer followed by a softmax to produce the output probability for each class.
    
\subsection{ncRNA}
RNA, together with DNA, is one of the fundamental carriers of genetic information. While the main function of RNA is in the production of proteins from instructions present in DNA code, RNA has been shown to also carry other important biological functions. In particular, recent findings on cancer research have shifted the attention from 
protein-coding RNAs to non-coding RNAs, as principal effectors and regulators 
of tumorigenesis and cancer development \cite{He2019, Wu2018, Tian2017, Yang2017}.  Moreover, certain RNAs have been shown to impact gene expression through fulfilling roles encompassing sensory and scaffolding capacities at various stages of the gene regulation process \cite{Mercer2013}. We refer to this functional RNA, which is transcribed from DNA but not translated into proteins, as non-protein coding RNA (ncRNA).

At its most basic form, RNA is a sequence of four types of nucleotides: \textit{adenine} (A), \textit{guanine} (G), \textit{cytosine} (C) and \textit{uracil} (U). The sequence of nucleotides forms what is commonly called the RNA \textit{primary structure}. 
% However, it is the 3D structure (or folding) of the RNA molecule which determines its function. This 3D structure is called tertiary structure and is generated by hydrogen bonds between complimentary base pairs. 
However, it is the folding of the RNA sequence into its \textit{secondary structure} which is more related to its function. The folding is generated by hydrogen bonds between complimentary base pairs. 
% The graph representing the folding is referred to as RNA secondary structure.
The most common occurring base pairs are A-U and G-C, also called Watson-Crick base pairs. However, RNA sometimes presents hydrogen bonds between different bases. All base pairs which do not follow Watson-Crick rules are called Wobble base pairs. Figure \ref{fig:rna_graph} shows the graph representing the folding of a short ncRNA sequence. We can observe both Watson-Crick pairs and Wobble pairs (G-U). Black edges represent phosphodiester bonds (between adjacent bases in the original sequence) while red edges represent hydrogen bonds.

\subsection{ncRNA Classification}
A wide variety of different classes, or families, of ncRNA have been identified, which differ by function and structure. Since the identification of drugs targeting the regulatory circuits of ncRNA depends on knowing its family, there has been an increasing interest in the development of methods for ncRNA classification. More traditional methods, such as \textit{RNA-CODE} \cite{RNA-code}, are based on alignment strategies. Other methods, such as \textit{RNAcon} \cite{Panwar2014} and \textit{GraPPLe} \cite{Childs2009}, use standard machine learning classifiers on manually extracted graph properties of the RNA secondary structure. These approaches have shown that graph properties (both local and global) reflect the functional information of different classes of RNAs and are therefore informative for the classification. 

More recently, \textit{nRC} \cite{Fiannaca2017} uses a convolutional neural network on graph features extracted using MoSS \cite{Borgelt:2005:MPM:1133905.1133908}, which finds frequent local sub-structures in a set of graphs. In particular, MoSS is used to extract up to 6483 binary features for each input graph, where each feature represents the presence or absence of a particular sub-structure. To the best of our knowledge, \textit{nRC} represents the state-of-the-art approach for ncRNA classification.

\subsection{Graph Convolutional Networks}
Deep learning has recently had a remarkable impact on multiple domains, including  natural language processing and computer vision \cite{DBLP:journals/nature/LeCunBH15}. However, most of popular deep neural models, such as convolutional neural networks (CNNs) \cite{726791}, only work on grid-structured (Euclidean) data, and are not directly applicable to graphs.
For this reason, \textit{nRC} \cite{Fiannaca2017} first extracts features from the ncRNA graphs before applying a CNN. 

Recently, there has been growing interest in extending deep learning techniques to non-Euclidean data, including graphs \cite{7974879}. Several models for deep learning on graphs have been developed in the past few years, including graph convolution \cite{DBLP:conf/iclr/KipfW17}, graph attention \cite{DBLP:conf/iclr/VelickovicCCRLB18}, mixture models \cite{DBLP:journals/corr/MontiBMRSB16} and neural message passing \cite{pmlr-v70-gilmer17a}.

\subsection{Our Contribution}
We are the first to apply graph convolutional networks on RNA folding data, achieving state-of-the-art results on the task of ncRNA classification with an accuracy of \textit{85.73\%} and an F1-score of \textit{85.61\%} over 13 classes. Our model is aware of different bond types and uses attention to aggregate information from the most important nodes for the final classification task. Moreover, since it learns directly from the RNA graphs, it removes the need for for manual features extraction.

\section{Background}
Most graph convolutional networks model can be interpreted as following a standard framework of \textit{message passing} \cite{pmlr-v70-gilmer17a}.  In particular, at each layer, the features of a node are updated by aggregating messages from its neighbors. Given a graph $G$, with node features $x_v$ and edge features $e_{vw}$, the update at layer $t+1$ takes the form:

\begin{align}
    m_v^{t+1} &= \sum_{w \in N(v)} M_t(x_v^t, x_w^t, e_{vw}) \\
    x^{t+1}_v &= U_t(x^t_v, m_v^{t+1})
\end{align}

where $M_t$ is a learnable function which computes the message from node $w$ to node $v$, $N(v)$ represents the neighbors of $v$ in the graph, $m_v^{t+1}$ represents the aggregation of all incoming messages for node $v$, and $U_t$ is a learnable function which updates the features for node $v$ given its previous features and the incoming aggregated message.

In graph classification problems, it is also necessary to produce a global graph representation by aggregating the final features for all nodes. This operation is called \textit{global pooling} and can be defined as:

\begin{equation}
    \hat{y} = R({x^T_v | v \in G})
\end{equation}

where $R$ is a learnable function which is permutation invariant with respect to the order of nodes, and $x^T_v$ represents the features of node $v$ after the last convolutional layer. 

% \begin{table*}[t]
%     \centering
%     \begin{tabular}{| c | clllll |}
%     \hline
%     \textit{Model} & \textit{Accuracy} & \textit{Sensitivity} & \textit{Specificity} & \textit{Precision} & \textit{F-score} & \textit{MCC} \\ \hline
%     \textit{RNACon} & 37.17\% & 37.17\% & 96.26\% & 45.84\% & 41.05\% & 33.43\% \\ \hline
%     \textit{nRC} & 81.04\% & 81.04\% & 98.42\% & 82.11\% & 81.57\% & 79.46\% \\  \hline
%     \textit{RNAGCN (ours))} & \textbf{85.29}\% & & & \textbf{87.82}\% & \textbf{86.30}\% & \textbf{84.07}\% \\ 
%     \hline
%     \end{tabular}
%     \caption{Summary of results on the independent test dataset with 12 classes. Results for RNACon and nRC are taken from \cite{Fiannaca2017}.}
%     \label{tab:results_12_classes}
% \end{table*}

\section{Proposed Method}
Our model is shown in figure \ref{fig:model}. It takes as input a graph corresponding to a folded ncRNA sequence. Mathematically, the ncRNA classification task takes the form of a prediction on a graph $G$ with node features $x_v$ and edge features $e_{vw}$. In particular, $x_v$ is just a one-hot representation of the nucleotide of node $v$, and $e_{vw}$ is a one-hot representation of the edge type (either hydrogen bond or phosphodiester bond).

The model consists of one embedding layer which maps each nucleotide to a continuous vector representation, followed by a sequence of graph convolutional layers. In particular, we use a layer similar to the one used in \cite{pmlr-v70-gilmer17a}, which is able to propagate information differently based on the edge type. Our convolutional layers take the form:

\begin{equation}
    x_v' = Wx_v + \sum_{w \in N(v)} A(e_{vw})x_w 
\end{equation}

where A is a 2-layer MLP with Leaky-ReLU as non linearity, which produces a projection matrix from edge features $e_{vw}$. Since our edge features $e_{vw}$ are one-hot encodings, this amounts to learning a different projection matrix for each edge type, allowing the model to spread information differently based on the bond between two nodes. Lastly, $W$ is a matrix of learnable weights.

After the last convolutional layer, a global pooling mechanism is used to obtain a single representation for the whole graph. In particular, we use the Set2Set model \cite{DBLP:journals/corr/VinyalsBK15}, which is a permutation invariant global pooling operator based on iterative content-based attention:

\begin{align}
q_t &= LSTM(q^*_{t-1})\\
\tilde{\alpha}_{v,t} &= x_v^\intercal q_t\\
\alpha_{v,t} &= \frac{exp(\tilde{\alpha}_{v,t})}{\sum_{w=1}^N exp(\tilde{\alpha}_{w,t})}\\
r_t &= \sum_{v=1}^N \alpha_{v,t}x_v \\
q^*_t &= q_t \Vert r_t
\end{align}

where $q_0$ is the zero vector, and $q_T^*$ is the final representation of the graph. At each step $t$, the output of the LSTM is used to compute attention scores $\alpha_{v,t}$ over all nodes. The new input to the LSTM is the concatenation of the old input with a weighted average of the nodes features, where the weights are given by the attention scores. We use $T=10$ steps and 1 layer for the LSTM. The size of the hidden state is the same as the number of output features from the last convolutional layer. 

A fully connected layer with softmax activation is finally used to produce the probabilities for each of the thirteen ncRNA classes. 

Each convolutional layer is followed by batch norm \cite{DBLP:journals/corr/IoffeS15}, before using the Leaky-ReLU activation function. Dropout \cite{Srivastava:2014:DSW:2627435.2670313} has also been used to regularize the model.

\section{Experiments}

\subsection{Dataset}
% \begin{table}
%     \centering
%     \begin{tabular}{| c | clll |}
%     \hline
%     \textit{Dataset} & \textit{\#Graphs} & \textit{Avg. \#Nodes} & \textit{Avg. \#Edges} & \textit{\#Classes}  \\ \hline
%     \textit{train} & 5670 & 162.02 & 210.46 & 13 \\  
%     \textit{val} & 650 & 163.30 & 212.12 & 13 \\ 
%     \textit{test} & 2600 & 149.15 & 193.25 & 13 \\
%     \hline
%     \end{tabular}
%     \caption{Summary of datasets statistics.}
%     \label{tab:data_analysis}
% \end{table}

\begin{table}
    \centering
    \begin{tabular}{| c | clll |}
    \hline
    \textit{Dataset} & \textit{\#Graphs} & \textit{Avg. \#Nodes} & \textit{Avg. \#Edges} & \textit{\#Classes}  \\ \hline
    \textit{train} & 5670 & 162.02 & 210.46 & 13 \\  
    \textit{val} & 650 & 163.30 & 212.12 & 13 \\ 
    \textit{test13} & 2600 & 149.15 & 193.25 & 13 \\
    \textit{test12} & 2400 & 147.52 & 191.13 & 12 \\
    \hline
    \end{tabular}
    \caption{Summary of datasets statistics.}
    \label{tab:data_analysis}
\end{table}

\renewcommand{\arraystretch}{1.5}
\begin{table}
    \centering
    \begin{tabular}{| c | c |}
    \hline
    \textit{Metric} & \textit{Formula}  \\ \hline
    Accuracy & $\frac{TP+TN}{TP+TN+FP+FN}$ \\  
    Sensitivity & $\frac{TP}{TP+FN}$ \\ 
    Specificity & $\frac{TN}{TN+FP}$ \\
    Precision & $\frac{TP}{TP+FP}$ \\
    F1-Score & $\frac{2*TP}{2*TP+FP+FN}$ \\
    MCC & $\frac{TP*TN-FP*FN}{\sqrt{(TF+FP)(TP+FN)(TN+FP)(TN+FN)}}$ \\
    \hline
    \end{tabular}
    \caption{Definition of the metrics used for the evaluation of the models.}
    \label{tab:metrics}
\end{table}
\renewcommand{\arraystretch}{1}

% \begin{table*}[t]
%     \centering
%     \begin{tabular}{| c | clllll |}
%     \hline
%     \textit{Model} & \textit{Accuracy} & \textit{Sensitivity} & \textit{Specificity} & \textit{Precision} & \textit{F1-score} & \textit{MCC} \\ \hline
%     \textit{nRC} & 81.81\% & 81.81\% & 98.48\% & 81.50\% & 81.66\% & 80.29\% \\  \hline
%     \textit{RNAGCN (ours)} & \textbf{85.73}\% & \textbf{86.09}\% & \textbf{98.82}\% & \textbf{86.09}\% & \textbf{85.61}\% & \textbf{84.59}\% \\ 
%     \hline
%     \end{tabular}
%     \caption{Summary of results on the independent test dataset with 13 classes. Results for nRC are taken from \cite{Fiannaca2017}.}
%     \label{tab:results_13_classes}
% \end{table*}

\begin{table*}[t]
    \centering
    \begin{tabular}{| c | c | clllll |}
    \hline
    \textit{Model} & \textit{Dataset} & \textit{Accuracy} & \textit{Sensitivity} & \textit{Specificity} & \textit{Precision} & \textit{F1-score} & \textit{MCC} \\ \hline
    \textit{nRC} & \textit{test13} & 81.81\% & 81.81\% & 98.48\% & 81.50\% & 81.66\% & 80.29\% \\  \hline
    \textit{RNAGCN (ours)} & \textit{test13} & \textbf{85.73}\% & \textbf{86.09}\% & \textbf{98.82}\% & \textbf{86.09}\% & \textbf{85.61}\% & \textbf{84.59}\% \\ 
    \hline \hline
    \textit{RNACon} & \textit{test12} & 37.17\% & 37.17\% & 96.26\% & 45.84\% & 41.05\% & 33.43\% \\ \hline
    \textit{nRC} & \textit{test12} & 81.04\% & 81.04\% & 98.42\% & 82.11\% & 81.57\% & 79.46\% \\  \hline
    \textit{RNAGCN (ours))} & \textit{test12} & \textbf{85.29}\% & \textbf{81.06}\% & \textbf{98.78}\% & \textbf{87.82}\% & \textbf{86.30}\% & \textbf{84.07}\% \\ \hline
    \end{tabular} 
    \caption{Summary of results on the two independent test datasets with 13 and 12 classes respectively. Results for \textit{nRC} and \textit{RNACon} are taken from \cite{Fiannaca2017}.}
    \label{tab:results_13_classes}
\end{table*}

We used the datasets introduced in \cite{Fiannaca2017}, which consist of a training dataset of 6320 ncRNA sequences and a test dataset of 2600 sequences. Both dataset contain sequences from 13 different ncRNA classes: \textit{miRNA}, \textit{5S rRNA}, \textit{5.8S rRNA}, \textit{ribozymes}, \textit{CD-box}, \textit{HACA-box}, \textit{scaRNA}, \textit{tRNA}, \textit{Intron gpI}, \textit{Intron gpII}, \textit{IRES}, \textit{leader} and \textit{riboswitch}. While the test dataset is perfectly balanced, the training dataset contains only 320 sequences from the \textit{IRES} class, compared to 500 sequences for all other classes.

In line with \cite{Fiannaca2017}, we also report results on a different test dataset, obtained by removing all sequences belonging to the \textit{scaRNA} class from the original test dataset. This allows for a comparison with \textit{RNACon} \cite{Panwar2014}, which was not trained on the \textit{scaRNA} class. We refer to the original test dataset with 13 classes as \textit{test13} and to the reduced test dataset with 12 classes as \textit{test12}.

In order to tune our model, we further split the original training dataset in two: one validation set with 650 sequences (50 from each class) and a training set with the remaining 5670 sequences \footnote{The code and the datasets are available at [Link Available Upon Acceptance]}. The statistics of the final splits are shown in table \ref{tab:data_analysis}. 

For each sequence, we generate the corresponding folding graph using the ViennaRNA \cite{Hofacker2003} package. 

\subsection{Experimental Setting}
The hyperameters of the model have been tuned on the held-out validation set using early stopping with a patience of 30 epochs. For the optimization of the model we used Adam \cite{DBLP:journals/corr/KingmaB14} with a learning rate of 0.0004. 

% hidden dimension of the convolutional layers = [40, 60, 80, 100]
% dropout rate = [0.1, 0.2, 0.5]
% number of conv layers = [3,4,5,6,7]
% global pooling mechanism = [sum, set2set]
% residuals = [False, True]

The best performing model on the validation set consists of 5 convolutional layers of dimension 80, the set2set model for global pooling, and uses a dropout rate of 0.1 for regularization.

For the evaluation of the model, we use the same metrics as in \cite{Fiannaca2017}, which we define in table \ref{tab:metrics}.

\subsection{Results}
We first tested our method on the independent test set with 13 classes (\textit{test13}). The results are shown in the top part of table \ref{tab:results_13_classes}. While \textit{nRC} obtains an accuracy of \textit{81.81\%}, our model outperforms it with an accuracy of \textit{85.73\%}. We also observe similar improvements on all other metrics.

The bottom part of table \ref{tab:results_13_classes} shows instead the results on the test dataset with only 12 classes (\textit{test12}). Our model outperforms both \textit{RNACon} and \textit{nRC} on all metrics. 

% We tested our method on the independent test set used by \cite{Fiannaca2017} and containing 2600 sequences belonging to 13 classes. The results are shown in table \ref{tab:results_13_classes}. While \textit{nRC} obtains an accuracy of \textit{81.81\%}, our model outperforms it with an accuracy of \textit{85.73\%}. We also observe similar improvements on all other metrics.

\section{Conclusion}
We have presented \textit{RNAGCN}, the first successful application of graph convolutional networks to RNA folding data, which achieves state-of-the-art results on the challenging task of ncRNA classification. Our model combines edge-aware convolutions and an attention-based pooling mechanism. With respect to existing approaches, our model comes with the additional benefit of being trained end-to-end and removing the need for manual feature extraction from the graph.

% \begin{acks}
% \end{acks}

\bibliographystyle{ACM-Reference-Format}
\bibliography{references}

% \appendix

\end{document}